\documentclass[twocolumn,aps,pra,superscriptaddress,showpacs,tightenlines]{revtex4-1}
\usepackage{amsmath}
\usepackage{amsfonts}
\usepackage{graphicx}
\usepackage{epsfig}
\usepackage{color}
\usepackage{hyperref}
\hypersetup{
    colorlinks=true, 
    linktoc=all,     
    linkcolor=blue,  
    citecolor=blue}

\begin{document}

\title{Cavity-assisted Dynamical Quantum Phase Transition at Bifurcation Points}

\author{Lin Tian}
\email{ltian@ucmerced.edu}
\affiliation{School of Natural Sciences, University of California, Merced, California 95343, USA}

\begin{abstract}
Coupling a quantum many-body system to a cavity can create bifurcation points in its phase diagram, where the ground state makes sudden switchings between different phases. Here we study the dynamical quantum phase transition of a transverse field Ising model coupled to a cavity. We show that an infinitesimal quench of the cavity driving at the bifurcation points induces gradual evolution of the Ising model to pass across the quantum critical point and excites quasiparticles. Meanwhile, when the driving is slowly ramped through the bifurcation points, the adiabaticity of the evolution and the number of quasiparticle excitations are strongly affected by cavity-induced nonlinearity. Introducing and manipulating cavity-induced nonlinearity hence provide an effective approach to control the dynamics and the adiabaticity of adiabatic quantum processes. Our model can be implemented with superconducting quantum circuits. 
\end{abstract}
\maketitle

\section{Introduction\label{sec:intro}}
Stimulated by recent experimental advances in manipulating atomic and solid-state quantum systems~\cite{LangenAnnuRevCMP2015, DevoretScience2013}, out-of-equilibrium many-body phenomena have been intensively studied~\cite{DziarmagaAdvPhys2010, PolkovnikovRMP2011, DelCampoIJMPA2014, EisertNPhys2015}. One intriguing effect is the dynamical quantum phase transition of a many-body system swept across its quantum critical point. This effect was first studied using scaling laws in the seminal works of \cite{ZurekPRL2005, PolkovnikovPRB2005}. It was predicted using the Kibble-Zurek mechanism that the adiabaticity of a slowly-ramped system will be broken near the critical point, which is accompanied by the production of quasiparticles~\cite{kzm1, kzm2, kzm3, kzm4}. Various approaches to improve the adiabaticity in such dynamical processes have been developed~\cite{BarankovPRL2008, SenPRL2008, ChandranPRB2012, CanevaPRL2009, NWuPRB2015, delCampo2012, Damski2014, DeffnerPRX2014, RezakhaniPRL2009}.  

Cavity or circuit quantum electrodynamics (QED) is a powerful approach to manipulate and probe the states of quantum systems, as demonstrated in earlier experiments on single qubit~\cite{cavityQED, circuitQED1, circuitQED2}. For a cavity coupled to a large number of atoms or qubits, collective effects such as bistability and Dicke superradiance have been studied~\cite{RitschRMP2013, EmaryPRE2003, MaschlerPRL2005, LarsonPRL2008, WChenPRA2009, TianPRL2010}. In recent experiments, it was shown that multiple superconducting qubits can be coupled simultaneously to a superconducting transmission line resonator~\cite{FinkPRL2009, MachaNatComm2014}. Dynamical phase transition has been observed for a collection of atoms coupled to a dissipative cavity~\cite{KlinderPNAS2015}. Furthermore, arrays of optical or microwave cavities can be connected to form Jaynes-Cummings lattices to emulate quantum phase transitions of cavity polaritons~\cite{HouckNPhys2012, KSeoPRB2015}. Given these experimental progresses, dynamical effects of many-body models coupled to a cavity can be studied in various physical systems. 

The coupling between a many-body model and a cavity induces nonlinearity in the coupled system, which can result in a bistable regime in the phase diagram~\cite{MaschlerPRL2005, LarsonPRL2008, WChenPRA2009}. At the bifurcation points in the bistable regime, the many-body system makes abrupt switchings between different many-body phases. These switchings are accompanied by finite changes in system parameters and energies, resembling a first-order phase transition~\cite{TianPRL2010}. How the switchings at the bifurcation points occur is an open question. Answering this question will help us understand the nonequilibrium dynamics in cavity-coupled systems. Here we study dynamical quantum phase transitions of a transverse field Ising model (TFIM)~\cite{SachdevBook1999} embedded in a cavity. Using a time-dependent Bogoliubov method~\cite{DziarmagaPRL2005}, we calculate the time evolution of this system undergoing sudden quench or gradual ramping of the cavity driving through the bifurcation points. Our results not only explain the dynamics of the sudden phase switchings, but also show that the dynamics near the critical point and the quasiparticle excitations are strongly affected by cavity-induced nonlinearity. Because bifurcation points can be engineered in a broad variety of models that are coupled to a cavity by choosing appropriate coupling operators, this study can be extended to other many-body systems with various forms of coupling. Our findings hence indicate that the adiabaticity of dynamical quantum processes can be controlled by introducing and adjusting cavity-induced nonlinearity, which can be utilized to develop high-fidelity schemes for adiabatic quantum computing or combinatorial optimization problems~\cite{FarhiScience2001, Santoro2002, Boixo2014}. Furthermore, the model studied here can be realized with superconducting quantum circuits. The TFIM can be emulated by arrays of superconducting qubits~\cite{Levitov2001, YDWangPRB2007, TianPRL2010, MarquardtPRL2013, GellerPRA2013, LDuPRA2015}, and the cavity can be mimicked by superconducting resonators. 

The paper is organized as follows. In Sec.~\ref{sec:system}, we introduce the coupled system of the emulator and the cavity, and describe the mean-field approach for studying this system. We then study the bifurcation points and the switchings of the many-body phases at the bifurcation points in Sec.~\ref{sec:bifurcation}. The time-dependent Bogoliubov method for studying the dynamics of this system is described in this section as well. In Sec.~\ref{sec:suddenquench}, we study the dynamics of this system undergoing sudden quench of the cavity driving near the bifurcation points. In Sec.~\ref{sec:slowramping}, we present our calculation for the dynamics of this system when the cavity driving is slowly ramped through the bifurcation points. Schemes for measuring the number of quasiparticles excited during the dynamical phase transition are given in Sec.~\ref{sec:measurement}. The effects of qubit decoherence and the validity of the mean-field approximation are discussed in Sec.~\ref{sec:discussions}. Finally, conclusions are given in Sec.~\ref{sec:conclusions}.

\section{System\label{sec:system}}
Consider a one-dimensional TFIM coupled to a cavity. The Hamiltonian of the TFIM is ($\hbar=1$)
\begin{equation}
H_{s}=-B_{x}\sum_{i}\hat{\sigma}_{xi}-J_{0}\sum_{i}\hat{\sigma}_{zi}\hat{\sigma}_{z(i+1)},\label{Hqim}
\end{equation}
where $\hat{\sigma}_{xi}$ and $\hat{\sigma}_{zi}$ are the Pauli matrices for a spin-$1/2$ particle at site $i$, $B_{x}$ is a uniform transverse field along the $x$-axis, and $J_{0}$ is a ferromagnetic coupling between neighboring spins. The ground state of this model is in a ferromagnetic (paramagnetic) phase for weak (strong) transverse field with $B_{x}<J_{0}$ ($B_{x}>J_{0}$) with a continuous quantum phase transition occurring at the critical point $B_{x}=J_{0}$~\cite{SachdevBook1999}. We set $J_{0}=1$ throughout this paper for convenience of discussion. The TFIM is exactly solvable using the Jordan-Wigner transformation that converts Ising spins to fermionic particles. The ground state can be written as $|G\rangle =\prod_{k>0}\left|G_{k}\right\rangle$ with $|G_{k}\rangle =(u_{k}+\textrm{i}v_{k}\hat{c}_{k}^{\dag}\hat{c}_{-k}^{\dag})|0_{k,-k}\rangle$, where $\hat{c}_{k}$ ($\hat{c}_{k}^{\dag}$) is the annihilation (creation) operator of a fermionic particle with quasimomentum $k$, $|0_{k,-k}\rangle$ is the vacuum state of the $(k,-k)$ modes, $u_{k}=\cos\theta_{k}$, and $v_{k}=\sin\theta_{k}$ as defined in Eqns.~(\ref{eq:theta1}) and (\ref{eq:theta2}) in Appendix~\ref{sec:qising}. In the subspace of zero or even number of excitations, $k=\pm(2m-1)\pi/N$ with $m=1,\cdots, N/2$~\cite{DziarmagaPRL2005}. The Hamiltonian of the cavity is $H_{c}=H_{c0}+H_{\kappa}$. The Hamiltonian
\begin{equation}
H_{c0}=-\Delta_{c}\hat{a}^{\dagger}\hat{a}-\epsilon(t)(\hat{a}+\hat{a}^{\dagger})\label{eq:Hc0}
\end{equation}
describes a cavity driven by an external field with amplitude $\epsilon(t)$, and $H_{\kappa}$ describes the cavity-bath coupling. Here $\hat{a}$ ($\hat{a}^{\dag}$) is the annihilation (creation) operator of the cavity, $\Delta_{c}$ is the cavity detuning relative to the driving frequency, and the cavity is dissipative with damping rate $\kappa$. We assume the coupling between the TFIM and the cavity as
\begin{equation}
H_{int}= g\left(\hat{a}+\hat{a}^{\dag}\right)\sum_{i}\hat{\sigma}_{xi},\label{eq:Hint}
\end{equation}
which can be realized in various physical systems, such as the circuit QED system~\cite{circuitQED1, circuitQED2}. With this coupling, the spin operators shift the cavity displacement $(\hat{a}+\hat{a}^{\dag})$, and the cavity displacement modifies the transverse field of the TFIM. Note that other forms of coupling can also generate similar nonlinear effects as discussed in our previous work \cite{TianPRL2010}. The total Hamiltonian of this system is then $H_{t}=H_{s}+H_{int}+H_{c}$.

Below we use a mean-field approach to treat the interaction between the TFIM and the cavity mode: $(\hat{a}+\hat{a}^{\dag})\sum_{i}\hat{\sigma}_{xi}\approx(\hat{a}+\hat{a}^{\dag})X+x_{a}\sum_{i}\hat{\sigma}_{xi}-x_{a}X$, where $X=\langle\sum_{i}\hat{\sigma}_{xi}\rangle$ is the average of the spin operator $\sum_{i}\hat{\sigma}_{xi}$ and $x_{a}=\langle (\hat{a}+\hat{a}^{\dag})\rangle $ is the average of the cavity displacement operator. With this approach, the total Hamiltonian can be decomposed into $H_{t}\approx\widetilde{H}_{s}+\widetilde{H}_{c}$, where $\widetilde{H}_{s}$ is the Hamiltonian of the TFIM under an effective transverse field $\widetilde{B}_{x}=B_{x}-gx_{a}$ and $\widetilde{H}_{c}$ is for the cavity under a modified driving field $\widetilde{\epsilon}=\epsilon-gX$. The stationary state of this system can be written as a product of the cavity state and the TFIM state. The effective Hamiltonian of the TFIM (cavity) hence depends on the operator average of the cavity mode (Ising spins). The cavity acts as a knob that modifies a parameter of the TFIM, and at the same time, it also depends on the state of the TFIM. 

\begin{figure}
\includegraphics[clip,width=\columnwidth]{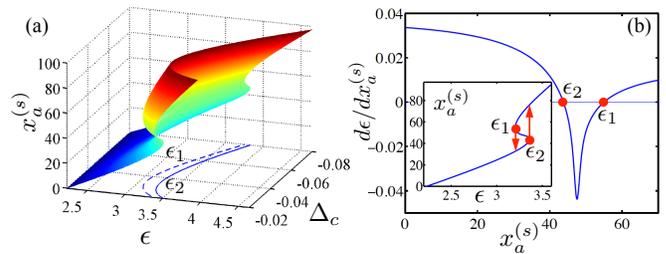}
\caption{(Color online) (a) The stationary cavity displacement $x_{a}^{(s)}$ vs. driving amplitude $\epsilon$ and detuning $\Delta_{c}$. Solid (dashed) line: the bifurcation point $\epsilon_{2}$ ($\epsilon_{1}$) vs. $\Delta_{c}$. (b) The derivative $d\epsilon/x_{a}^{(s)}$ vs. $x_{a}^{(s)}$ at detuning $\Delta_{c}=-0.05$. Inset: $x_{a}^{(s)}$ vs. $\epsilon$. Other parameters are $J_{0}=1$, $B_{x}=1.95$, $g=0.02$ and $\kappa=0.07$, all in arbitrary units; and the TFIM in our calculation has $N=120$ sites.}
\label{fig1}
\end{figure}
\section{Bifurcation and phase switching\label{sec:bifurcation}}
The Langevin equation for the average of the cavity operator $\hat{a}$ is~\cite{WallsMilburnBook}
\begin{equation}
\frac{d}{dt}\left\langle \hat{a}\right\rangle  = \textrm{i}\Delta_{c}\left\langle \hat{a}\right\rangle -\frac{\kappa}{2}\left\langle \hat{a}\right\rangle +\textrm{i}\left(\epsilon-gX\right),\label{eq:dat}
\end{equation}
which depends on the average $X=\langle\sum_{i}\hat{\sigma}_{xi}\rangle$ of the spin operator $\sum_{i}\hat{\sigma}_{xi}$. For the stationary state, $d\langle \hat{a}\rangle/dt=0$, and the cavity displacement can be derived as
\begin{equation}
x_{a}^{(s)} = \frac{2\Delta_{c}\left(-\epsilon+gX^{(s)}\right)}{\Delta_{c}^{2}+\kappa^{2}/4}.\label{eq:avexa}
\end{equation}
In the stationary state, the TFIM is in its ground state under the effective transverse field $\widetilde{B}_{x}=B_{x}-gx_{a}^{(s)}$, and $X^{(s)}$ is the ground-state average of the operator $\sum_{i}\hat{\sigma}_{xi}$. The operator average $X^{(s)}$ has a nonlinear dependence on the stationary displacement $x_{a}^{(s)}$ through its dependence on the effective transverse field $\widetilde{B}_{x}$, which yields a bistable regime in this system. Details of the nonlinear dependence and the bistability are given in Appendix~\ref{sec:stability}. At given driving amplitude $\epsilon$ and detuning $\Delta_{c}$ in the bistable regime, $x_{a}^{(s)}$ has three stationary solutions as shown in Fig.~\ref{fig1}(a). The stability condition~\cite{DeJesusPRA1987} for these solutions is $d\epsilon/dx_{a}^{(s)}>0$, as obtained in Appendix~\ref{sec:stability}. From the inset of Fig.~\ref{fig1}(b), we see that the largest and the smallest solutions are stable, whereas the intermediate solution is unstable. The bifurcation points divide the bistable regime from regimes with only one stable solution, and they satisfy the condition $d\epsilon/dx_{a}^{(s)}=0$ [indicated by red circles in Fig.~\ref{fig1}(b)]~\cite{DrazinBook1992}. In Fig.~\ref{fig1}(a), the bifurcation points $\epsilon_{1,2}$ are plotted versus $\Delta_{c}$, which form a phase diagram in the ($\epsilon$, $\Delta_{c}$) plane of this coupled system. The area between $\epsilon_{1}$ and $\epsilon_{2}$ corresponds to the bistable regime, where the TFIM can be in either the paramagnetic or the ferromagnetic phase, depending on the history of the dynamical evolution. 

At the bifurcation point $\epsilon_{2}$, the TFIM makes a transition from the paramagnetic phase to the ferromagnetic phase. An opposite switching occurs at $\epsilon_{1}$. Such switchings involve finite changes in $x_{a}^{(s)}$ and $\widetilde{B}_{x}$, resembling a first-order phase transition~\cite{TianPRL2010}. To study how the switchings occur, we use a time-dependent Bogoliubov method for the TFIM~\cite{DziarmagaPRL2005}. The dynamics of the TFIM is governed by the Schr\"{o}dinger equation $\textrm{i}d|\psi(t)\rangle/dt=\widetilde{H}_{s}(t)|\psi(t)\rangle $ for wave function $|\psi(t)\rangle$, where $\widetilde{H}_{s}$ contains a time-dependent effective transverse field $\widetilde{B}_{x}(t)$. Initially prepared in the ground state of an effective transverse field $\widetilde{B}_{x}(0)$, the wave function can always be written in the form of $|\psi(t)\rangle =\prod_{k>0}(U_{k}(t)+\textrm{i}V_{k}(t)\hat{c}_{k}^{\dagger}\hat{c}_{-k}^{\dag})|0\rangle$ with time-dependent coefficients $U_{k}(t)$ and $V_{k}(t)$, as discussed in Appendix~\ref{sec:qising}. Let $U_{k}=\bar{U}_{k}e^{-\textrm{i}\varphi(t)}$ and $V_{k}=\bar{V}_{k}e^{-\textrm{i}\varphi(t)}$, with phase factor $\varphi(t)=\int_{0}^{t}dt'[-2B_{x}(t^{\prime})+\varepsilon_{k}(t')\cos[2\theta_{k}(t')]]$ and eigenenergy $\varepsilon_{k}(t)$ for the effective transverse field $\widetilde{B}_{x}(t)$. We have
\begin{equation}
\textrm{i}\frac{d}{dt}\left[\begin{array}{c}
\bar{U}_{k}\\
\bar{V}_{k}
\end{array}\right]=\left[\begin{array}{cc}
-\varepsilon_{k}\cos(2\theta_{k}) & -\varepsilon_{k}\sin(2\theta_{k})\\
-\varepsilon_{k}\sin(2\theta_{k}) & +\varepsilon_{k}\cos(2\theta_{k})
\end{array}\right]\left[\begin{array}{c}
\bar{U}_{k}\\
\bar{V}_{k}
\end{array}\right].\label{eq:dUV}
\end{equation}
The dynamic matrix on the right hand side of Eq.~(\ref{eq:dUV}) depends on the time-dependent cavity displacement $x_{a}(t)$, whose dynamics is governed by Eq.~(\ref{eq:dat}). By solving Eqns.~(\ref{eq:dat}) and (\ref{eq:dUV}) jointly, we obtain the dynamics of this system. Note that the wave function $|\psi(t)\rangle$ can also be expressed as $|\psi(t)\rangle=\prod_{k>0}[\alpha_{k}(t)+\textrm{i}\beta_{k}(t)\hat{\gamma}_{k}^{\dag}\hat{\gamma}_{-k}^{\dag}]|G_{k}\rangle$ in terms of the instantaneous ground state $|G_{k}\rangle$ and the instantaneous excited state $\textrm{i}\hat{\gamma}_{k}^{\dag}\hat{\gamma}_{-k}^{\dag}|G_{k}\rangle$ with probability amplitudes $\alpha_{k}$ and $\beta_{k}$, respectively. Expressions of $\alpha_{k}$ and $\beta_{k}$ are given in Appendix~\ref{sec:qising}.

\section{Dynamics upon sudden quench\label{sec:suddenquench}}
\begin{figure}
\includegraphics[clip,width=\columnwidth]{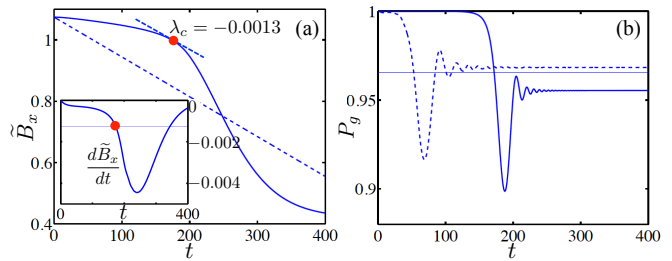}
\caption{(Color online) (a) The effective transverse field $\widetilde{B}_{x}$ and (b) the ground state probability $P_{g}$ vs. time $t$ after a sudden quench of the driving amplitude. Solid lines: results for the cavity-coupled TFIM; dashed lines: results for the linearly-ramped TFIM; thin solid line in (b): results using the Landau-Zener formula for the linearly-ramped TFIM. The inset of (a): $d\widetilde{B}_{x}/dt$ vs. time $t$. Red dots: the time $t_{c}$ for $\widetilde{B}_{x}=J_{0}$. The parameters are the same as these in Fig.~\ref{fig1}(b).}
\label{fig2} 
\end{figure}

The phase switchings at the bifurcation points can be triggered by an infinitesimally-small sudden quench of the cavity driving amplitude. Here we study the dynamics of this system after such a sudden quench. Let the system be prepared in a stationary state of the driving amplitude $\epsilon=\epsilon_{2}-\Delta\epsilon/2$, which is right below the bifurcation point $\epsilon_{2}$. We choose $\Delta\epsilon=0.01$ with $\Delta\epsilon\ll |\epsilon_{2}-\epsilon_{1}|$. At $B_{x}=1.95$ and $\Delta_{c}=-0.05$, $\epsilon_{2}=3.36$. With the stationary cavity displacement $x_{a}^{(s)}=43.78$ and the effective transverse field $\widetilde{B}_{x}=1.07$, the TFIM is in a paramagnetic phase. At time $t=0$, the driving field is tuned to $\epsilon=\epsilon_{2}+\Delta\epsilon/2$, which is right above the bifurcation point with the stationary cavity displacement $x_{a}^{(s)}=75.95$. At this driving amplitude, the stationary phase for the TFIM is ferromagnetic. After the quench, the system evolves towards the new stationary state. In Fig.~\ref{fig2}(a), we plot the evolution of the effective transverse field $\widetilde{B}_{x}(t)$ in a time interval $0<t<T$ with total evolution time $T=400$. The effective transverse field decreases gradually, and passed through the critical point $\widetilde{B}_{x}=1$ at time $t_{c}\sim160$. The system eventually reaches the new stationary state of the quenched driving amplitude with the final effective field $\widetilde{B}_{x}=0.43$. Similar quench behavior can be observed at the bifurcation point $\epsilon_{1}$. A tiny quench of the driving amplitude hence induces finite changes in the system parameters and causes a first-order-like phase transition in the TFIM, owing to the cavity-induced nonlinearity that creates a discontinuity in the stationary solutions at the bifurcation points.

When the effective transverse field approaches the critical point, the relaxation time of the TFIM diverges as $\sim 1/(2|\widetilde{B}_{x}-J_{0}|)$, and its relaxation becomes slower than the variation of the Hamiltonian. As explained by the Kibble-Zurek mechanism~\cite{kzm1, kzm2, kzm3, kzm4}, the evolution ceases to be adiabatic, and quasiparticles can be excited. In Fig.~\ref{fig2}(b), the probability $P_{g}$ for the TFIM in its instantaneous ground state is plotted for the time interval $0<t<T$. At $t\ll t_{c}$ far from the phase transition point, the TFIM evolves adiabatically and remains in its ground state with the probability $P_{g}\approx1$. Near time $t=t_{c}$, $P_{g}$ starts decreasing, and exhibits oscillatory behavior. The probability eventually stabilizes at the stationary value $P_{g}=0.955$. This indicates that quasiparticles are created in the TFIM during the evolution. It can be shown that during the entire evolution, $P_{g}\approx(1-|\beta_{k=\pi/N}|^{2})$, i.e., the quasiparticles are mainly excited in the $k=\pm\pi/N$ modes, which have the lowest excitation energy. This effect is also observed for switchings at the bifurcation point $\epsilon_{1}$ as well as for a wide range of transverse field $B_{x}$ and cavity detuning $\Delta_{c}$. 

For comparison, we simulate the dynamics of a simple TFIM under linearly-ramped transverse field $B_{x}(t)=\widetilde{B}_{x}(0)+\lambda_{c}t$~\cite{ZurekPRL2005, PolkovnikovPRB2005, DziarmagaPRL2005}. Here $\widetilde{B}_{x}(0)$ is the effective transverse field of the cavity-coupled TFIM at time $t=0$. The ramping rate $\lambda_{c}$ is chosen to be the slope of $\widetilde{B}_{x}(t)$ at time $t=t_{c}$, i.e., $\lambda_{c}=d\widetilde{B}_{x}/dt\vert_{t=t_{c}}$, as indicated in Fig.~\ref{fig2}(a). With our parameters, $\lambda_{c}=-0.0013$. The ground-state probability $P_{g}$ of this linearly-ramped model is plotted in Fig.~\ref{fig2}(b), and it shows similar behavior to that of the cavity-coupled TFIM. At time $t=T$, $P_{g}=0.967$ with only $1\%$ deviation from that of the cavity-coupled model. The onset of the probability decrease occurs earlier in this case than in the cavity-coupled model, as $B_{x}(t)$ reaches the critical point at an earlier time than that of $\widetilde{B}_{x}(t)$. In this model, we can estimate the quasiparticle excitations using the Landau-Zener formula~\cite{lz1, lz2}. As shown in \cite{DamskiPRL2005, MGongKZM2015}, the probability of the excitation of the $(k,-k)$ modes is
\begin{equation}
P_{k}^{(LZ)}=\exp{\left(-2\pi J_{0}^{2}k^{2}/\lambda_{c}\right)}.\label{eq:Pekn}
\end{equation}
The ground-state probability can then be written as $P_{g}^{(LZ)}=\prod_{k}(1-P_{k}^{(LZ)})$ with $P_{g}^{(LZ)}=0.966$, very close to the value of $P_{g}(T)$ from our numerical simulations. As shown in Fig.~\ref{figS3} in Appendix~\ref{sec:stability}, it provides accurate estimation of the ground-state probability for a wide range of transverse field $B_{x}$ and cavity detuning $\Delta_{c}$. Our study indicates that the adiabaticity of the cavity-coupled TFIM can be well characterized by a single parameter $\lambda_{c}$, which is strongly influenced by cavity-induced nonlinearity. 

\begin{figure}
\includegraphics[clip,width=\columnwidth]{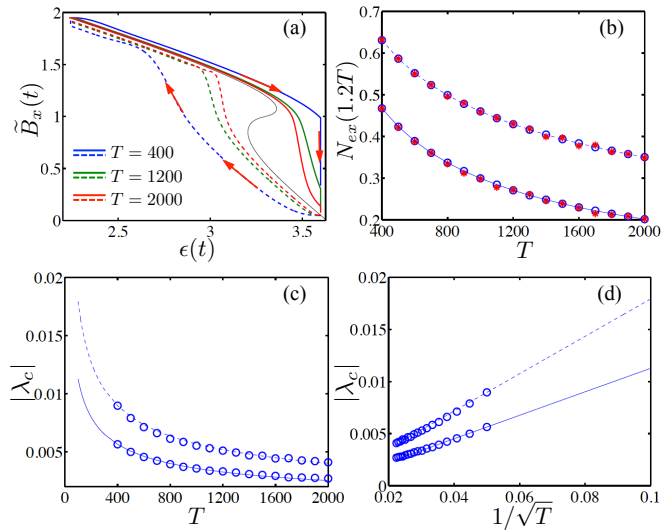}
\caption{(Color online) (a) The effective transverse field $\widetilde{B}_{x}$ vs. $\epsilon$ under slow ramping of the driving amplitude. Solid (dashed) lines: results for ramping up (down) at $T=400,\,1200,\,2000$ from top to bottom (bottom to top); thin solid line: stationary solution. (b) $N_{ex}$ at $t=1.2T$ vs. ramping time $T$. Blue circles and solid (dashed) line: results for cavity-coupled TFIM, ramping up (down); red stars: results using the Landau-Zener formula for linearly-ramped TFIM. (c) and (d) $|\lambda_{c}|$ vs. $T$ and $1/\sqrt{T}$. Blue circles: results from numerical simulation; solid (dashed) lines: results using the formula $|\lambda_{c}|=20|\lambda_{c}(T=400)|/\sqrt{T}$ for ramping up (down). The parameters are the same as that these in Fig.~\ref{fig1}(b).}
\label{fig3}
\end{figure}
\section{Dynamics under slow ramping\label{sec:slowramping}}
We now study dynamical phase transition in the cavity-coupled TFIM when the driving amplitude is gradually ramped across the bistable regime. Assume that the driving amplitude has the time dependence $\epsilon(t)=\epsilon_{0}+(\epsilon_{f}-\epsilon_{0})t/T$ in the time interval $0< t\le T$ and $\epsilon(t)=\epsilon_{f}$ in the time interval $T< t\le 1.2T$, where $\epsilon_{0}$ ($\epsilon_{f}$) is the initial (final) amplitude of the cavity driving and $T$ ($1.2T$) is the duration of the ramping (entire evolution). Here the driving amplitude is slowly ramped to reach the final amplitude $\epsilon_{f}$, and then it is parked at $\epsilon_{f}$ for a duration of $0.2T$ to allow the system to relax towards its new stationary state. Let $\epsilon_{0}=2.23$ for $B_{x}=1.95$ and $\Delta_{c}=-0.05$, which yields $x_{a}^{(s)}=0$ and $\widetilde{B}_{x}=B_{x}$. At $\epsilon_{f}=3.6$, the stationary transverse field becomes $\widetilde{B}_{x}=0.05$. With the time-dependent Bogoliubov method for the TFIM~\cite{DziarmagaPRL2005} and the Langevin equation (\ref{eq:dat}) for the cavity mode, we simulate the dynamics of this system. Figure~\ref{fig3}(a) shows the correspondence between the effective transverse field $\widetilde{B}_{x}(t)$ and the driving amplitude $\epsilon(t)$ during the evolution for several values of ramping times $T$. For the shortest ramping time $T=400$, $\widetilde{B}_{x}\approx 1$ by the end of the ramping at $t=T$, which barely crosses the critical point. The effective transverse field continues to decrease during the parking interval $T< t\le1.2T$, which produces a large vertical segment in the plot. For the longest ramping time $T=2000$, $\widetilde{B}_{x}$ nearly reaches its stationary value by the end of the ramping at $t=T$. In these evolutions, the TFIM remains in the paramagnetic phase when the driving amplitude reaches the bifurcation point $\epsilon_{2}$, and enters the ferromagnetic phase at a later time. A hysteresis loop that encloses the stationary solution can be observed when the driving amplitude is first ramped up to $\epsilon_{f}$ and then ramped back to $\epsilon_{0}$~\cite{DrazinBook1992}. For longer ramping time $T$, the hysteresis loop is closer to the stationary solution, as the system has more time to relax towards its stationary state. 

During the ramping, quasiparticles are created when the effective transverse field is at the vicinity of the critical point. In Fig.~\ref{fig3}(b), we plot the number of excited quasiparticle pairs $N_{ex}=\sum_{k>0}|\beta_{k}|^{2}$ at time $t=1.2T$ as a function of the ramping time $T$, with $\beta_{k}$ being the instantaneous excited-state  amplitude. It is shown that $N_{ex}$ decreases quickly with the ramping time $T$. With the Landau-Zener formula, we estimate the number of excitations for a linearly-ramped TFIM with ramping rate $\lambda_{c}$. This estimation agrees well with numerical results for the cavity-coupled TFIM, with a maximum deviation of only $0.007$. Hence, the quasiparticle excitation and the adiabaticity in the cavity-coupled model can be well characterized by the ramping rate $\lambda_{c}$. 

In Figs.~\ref{fig3}(c) and (d), the magnitude of the ramping rate $\vert\lambda_{c}\vert$ is plotted versus the ramping time $T$ and $1/\sqrt{T}$, respectively. We find that the dependence of $\lambda_{c}$ on the ramping time $T$ can be approximated by an empirical expression: $\lambda_{c}=20\lambda_{c}\vert_{T=400}/\sqrt{T}$, with a maximum discrepancy of $3.6\%$ from the exact numerical results. This inverse-square-root dependence of $\lambda_{c}$ on the ramping time $T$ is caused by the nonlinearity in the cavity-coupled TFIM. Note that the dependence of $\lambda_{c}$ on $T$ can be controlled by adjusting system parameters, such the coupling strength $g$ and the cavity detuning $\Delta_{c}$; and this dependence is different for different forms of qubit-cavity coupling. Our study hence shows that by introducing and manipulating cavity-induced nonlinearity, the adiabaticity of dynamical quantum processes in cavity-coupled many-body systems can be controlled. 

\section{Measurement of quasiparticle excitations\label{sec:measurement}}
During dynamical phase transitions, quasiparticles can be created in the TFIM when the effective transverse field $\widetilde{B}_{x}$ (or $B_{x}$ in the case of a bare TFIM not coupled to a cavity) reaches the vicinity of the quantum critical point. For an initial state that is the ground state under a given transverse field, the system remains in the even-parity subspace during the entire evolution, as discussed in Appendix~\ref{sec:qising}. As a result, quasiparticles are always generated in pairs, i.e., the $\hat{\gamma}_{k}$ and $\hat{\gamma}_{-k}$ particles are excited simultaneously. The operator for the number of quasiparticle pairs can be defined as $\hat{N}_{ex}=\sum_{k>0}\hat{\gamma}_{k}^{\dag}\hat{\gamma}_{k}$. Alternatively, $\hat{N}_{ex}=(1/2)\sum_{k}\hat{\gamma}_{k}^{\dag}\hat{\gamma}_{k}$, where the summation is taken over both positive and negative values of $k$. In terms of the probability amplitude $\beta_{k}$ of the excited states given by Eq.~(\ref{eq:psiab}) in Appendix~\ref{sec:qising}, the number of quasiparticle pairs is $N_{ex}=\langle \hat{N}_{ex}\rangle=\sum_{k>0}|\beta_{k}|^{2}$.

A key component in studying the dynamical quantum phase transition in this coupled system is the measurement of the quasiparticle excitations in the TFIM. Below we will show that the number of quasiparticle pairs can be measured in the limits of $\widetilde{B}_{x}\ll J_{0}$ (i.e., $\widetilde{B}_{x}\rightarrow 0$) and $\widetilde{B}_{x}\gg J_{0}$ (i.e., $J_{0}\rightarrow 0$) by directly measuring the qubit states in their physical basis. For intermediate values of the transverse field with the magnitude of $\widetilde{B}_{x}$ comparable to that of $J_{0}$, however, it is nontrivial to directly measure the quasiparticle excitations. In this work, the effective transverse field at the end of the dynamical evolutions is always far above or far below the critical point. For an effective transverse field far below (above) the critical point, we can first reduce the effective transverse field adiabatically to the limit of $\widetilde{B}_{x}\ll J_{0}$ ($\widetilde{B}_{x}\gg J_{0}$), which will not induce more quasiparticle excitations. The number of quasiparticle excitations can then be measured in these limits. 

In the limit of weak transverse field with $\widetilde{B}_{x}\ll J_{0}$ ($\widetilde{B}_{x}\rightarrow 0$), the eigenenergies are $\varepsilon_{k}=2J_{0}$ for all values of $k$. The Hamiltonian of the TFIM then becomes $H_{s}=2J_{0}\sum_{k}\hat{\gamma}_{k}^{\dag}\hat{\gamma}_{k}+E_{g}$ with ground-state energy $E_{g}=-J_{0}N$, where the summation is taken over all values of $k$. At the same time, Eq.~(\ref{Hqim}) shows that $H_{s}=-J_{0}\sum_{i}\hat{\sigma}_{zi}\hat{\sigma}_{z(i+1)}$ in this limit. The number of quasiparticle pairs can hence be written as
\begin{equation}
N_{ex}=\langle N-\sum_{i}\hat{\sigma}_{zi}\hat{\sigma}_{z(i+1)}\rangle/4,\label{eq:Nex}
\end{equation}
which is directly related to the number of spin flips in the $\hat{\sigma}_{z}$-basis in the TFIM. In the even-parity subspace, the ground state at $\widetilde{B}_{x}=0$ is $|G\rangle=(|\uparrow,\uparrow,\cdots,\uparrow\rangle+|\downarrow,\downarrow,\cdots,\downarrow\rangle)/\sqrt{2}$, where all spins are aligned with each other and $\langle\hat{\sigma}_{zi}\hat{\sigma}_{z(i+1)}\rangle\equiv1$. Hence $N_{ex}=0$ in the ground state. With one spin flip at, e.g., site $i_{f}$, $\langle\hat{\sigma}_{zi_{f}}\hat{\sigma}_{z(i_{f}+1)}\rangle=\langle\hat{\sigma}_{z(i_{f}-1)}\hat{\sigma}_{zi_{f}}\rangle=-1$. This yields $N_{ex}=1$, i.e., one pair of quasiparticles are excited. By measuring the qubits in their $\hat{\sigma}_{z}$-basis, the number of quasiparticle pairs can then be determined. 

In the opposite limit of $\widetilde{B}_{x}\gg J_{0}$ ($J_{0}\rightarrow 0$), the Hamiltonian of the TFIM can be written as $H_{s}=2B_{x}\sum_{k}\hat{\gamma}_{k}^{\dag}\hat{\gamma}_{k}-NB_{x}$ with ground-state energy $E_{g}=-NB_{x}$ and $\varepsilon_{k}=2B_{x}$ for all $k$. From Eq.~(\ref{Hqim}), the Hamiltonian has the form $H_{s}=-B_{x}\sum_{i}\hat{\sigma}_{xi}$. We then have
\begin{equation}
N_{ex}=\langle N-\sum_{i}\hat{\sigma}_{xi}\rangle/4,\label{eq:Nex}
\end{equation}
which is related to the spin alignment in the $\hat{\sigma}_{x}$-basis. The ground state in this limit is $|G\rangle=|\rightarrow,\rightarrow,\cdots,\rightarrow\rangle$ with all spins  aligned along the $x$-axis, which corresponds to $N_{ex}=0$. Note that for a state with a single spin flip in the $\hat{\sigma}_{x}$-basis, $N_{ex}=1/2$, corresponding to the excitation of a single quasiparticle. Such states belong to the odd-parity subspace and cannot be created in the evolutions studied in this work, when the TFIM is initially prepared in the ground state of a given effective transverse field. Hence the number of quasiparticle pairs can be determined by measuring the qubits in the $\hat{\sigma}_{x}$-basis.

\section{Discussions\label{sec:discussions}}
The cavity-coupled TFIM can be realized with superconducting systems. It was shown that superconducting qubits, such as flux qubits and transmons, can be connected in an array to form the one-dimensional TFIM~\cite{TianPRL2010, Levitov2001, YDWangPRB2007, MarquardtPRL2013, GellerPRA2013, LDuPRA2015}. The qubits can be coupled to a superconducting resonator that acts as the cavity mode. Moreover, the dynamical effects studied here can be demonstrated using finite-sized TFIM's that contain only a few qubits. In our analysis, we neglect the decoherence of the qubits, which could cause the quasiparticles excited during the evolution to decay. In order to reliably characterize the dynamical quantum phase transition and the scaling of the quasiparticle excitations, the duration of the evolution is required to be much shorter than the decay time caused by qubit decoherence. In superconducting qubits, decoherence times on the order of $100\mu\textrm{s.}$ have been observed in recent experiments~\cite{DevoretScience2013}. For a coupling strength of $J_{0}/2\pi=1\textrm{GHz}$, a duration of $T=2000$ corresponds to $318\,\textrm{ns.}$, much shorter than the decoherence time of the qubits. 

In addition, when treating the interaction between the qubits and the cavity mode, we use a mean-field approach to decompose the total Hamiltonian into separate parts for the TFIM and for the cavity mode. This approach is valid for cavity fields under strong driving~\cite{meanfield}. With our parameters, the cavity state can be viewed as a coherent state with large amplitude $\langle\hat{a}\rangle\sim50$ and large photon number $N_{a}\sim2500$. The ratio of the quantum fluctuations with respect to the cavity photon number is then $1/\sqrt{N_{a}}\sim0.02\ll 1$, and the fluctuations can hence be neglected.

\section{Conclusions\label{sec:conclusions}}
To conclude, we study the dynamical quantum phase transitions across the bifurcation points of a quantum many-body system coupled to a cavity mode. In particular, we calculate the time evolutions of a cavity-coupled TFIM undergoing sudden quench or slow ramping of the cavity driving. Our numerical simulations explain the mechanism of the phase switchings across the bifurcation points. Moreover, our results show that cavity-induced nonlinearity can strongly affect the adiabaticity of quantum dynamical processes in such many-body systems. Our findings can be exploited to improve the adiabaticity and fidelity of various adiabatic quantum processes, such as adiabatic quantum computing and combinatorial optimization protocols, via cavity-assisted schemes.

\begin{acknowledgements}
This work is supported by the National Science Foundation under Award No. 0956064. The author thanks the Institute of Physics, Chinese Academy of Sciences, for hospitality. 
\end{acknowledgements}

\appendix
\section{Properties of the transverse field Ising model\label{sec:qising}}
In this section, we present properties of a one-dimensional (1D) TFIM in detail, including quasiparticle excitations, ground state, time-dependent wave function, and operator averages.

\subsection{Quasiparticle excitations\label{ssec:quasi}}
The TFIM, also called the quantum Ising model, has been widely studied as a prototype for continuous quantum phase transition~\cite{SachdevBook1999}. Consider the Hamiltonian (\ref{Hqim}) for a 1D TFIM. We assume that the spin operators satisfy the periodic boundary condition: $\hat{\sigma}_{\alpha (N+1)}=\hat{\sigma}_{\alpha 1}$ with $\alpha=x,\,y,\,z$. This model is exactly solvable by applying the Jordan-Wigner transformation~\cite{JWT}
\begin{equation}
\hat{\sigma}_{zi} =  -(\hat{c}_{i}^{\dagger}+\hat{c}_{i})\prod_{j=1}^{i-1}(1-2\hat{c}_{j}^{\dagger}\hat{c}_{j}),\quad \hat{\sigma}_{xi}= (1-2\hat{c}_{i}^{\dagger}\hat{c}_{i}),\label{eq:JWT}
\end{equation}
where $\hat{c}_{i}$ ($\hat{c}_{i}^{\dag}$) is the annihilation (creation) operator of a fermionic particle at site $i$. This transformation converts the Ising spins in the 1D TFIM to spineless fermions. In the fermionic representation, the Hamiltonian (\ref{Hqim}) becomes 
\begin{equation}
H_{s} = -J_{0}\sum (\hat{c}_{i}^{\dagger}\hat{c}_{i+1}^{\dagger}+\hat{c}_{i}^{\dagger}\hat{c}_{i+1}+h.c.)+2B_{x}\sum\hat{c}_{i}^{\dagger}\hat{c}_{i}-NB_{x}.\label{eq:cfermion}
\end{equation}
The fermionic operators satisfy the anti-periodic boundary condition: $\hat{c}_{N+1}=-\hat{c}_{1}$, in the subspace of zero or even number of quasiparticle excitations~\cite{DziarmagaPRL2005}. Define the momentum-space operators $\hat{c}_{k}=\sum_{i}e^{-\textrm{i}k i}\hat{c}_{i}/\sqrt{N}$, where the wave vectors are $k=\pm\pi(2m-1)/N$ with $1\le m\le N/2$ and are odd multiples of $\pi/N$ under the anti-periodic boundary condition. In the momentum basis, the Hamiltonian becomes
\begin{eqnarray}
 H_{s} &=& \sum_{k>0}\left[2\left(B_{x}-J_{0}\cos k\right)\left(\hat{c}_{k}^{\dagger}\hat{c}_{k}+\hat{c}_{-k}^{\dagger}\hat{c}_{-k}\right)\right.\nonumber \\
 &&-\left. 2\textrm{i}J_{0}\sin k\left(\hat{c}_{k}^{\dagger}\hat{c}_{-k}^{\dagger}-\hat{c}_{-k}\hat{c}_{k}\right)\right]-NB_{x},\label{eq:cfermion-k}
\end{eqnarray}
where only $k$ and $(-k)$ modes are coupled to each other.

The quasiparticle modes are defined with the Bogoliubov transformation
\begin{subequations}
\begin{alignat}{2}
\hat{c}_{k} & =  u_{k}\hat{\gamma}_{k}+\textrm{i}v_{k}\hat{\gamma}_{-k}^{\dag},\label{eq:bt1}\\
\hat{c}_{-k}^{\dag} & =  \textrm{i}v_{k}\hat{\gamma}_{k}+u_{k}\hat{\gamma}_{-k}^{\dag},\label{eq:bt2}
\end{alignat}
\end{subequations}
where $\hat{\gamma}_{k}$ ($\hat{\gamma}_{k}^{\dag}$) is the annihilation (creation) operator of the quasiparticle with momentum $k$. The coefficients in this transformation satisfy the relations $u_{-k}=u_{k}$, $v_{-k}=-v_{k}$, and $u_{k}^{2}+v_{k}^{2}=1$. The quasiparticle operators satisfy the commutation relations: $[\hat{\gamma}_{k},\,\hat{\gamma}_{l}^{\dag}]_{+}=\delta_{kl}$ and $[\hat{\gamma}_{k},\,\hat{\gamma}_{l}]_{+}=0$, where $[\hat{A},\hat{B}]_{+}$ is the anti-commutator for operators $\hat{A}$ and $\hat{B}$. It can also be shown that
\begin{subequations}
\begin{alignat}{2}
\hat{\gamma}_{k} & =  u_{k}\hat{c}_{k}-\textrm{i}v_{k}\hat{c}_{-k}^{\dag},\label{eq:btrev1}\\
\hat{\gamma}_{-k}^{\dag} & =  -\textrm{i}v_{k}\hat{c}_{k}+u_{k}\hat{c}_{-k}^{\dag}.\label{eq:btrev2}
\end{alignat}
\end{subequations}
The coefficients can be written as $u_{k}=\cos\theta_{k}$ and $v_{k}=\sin\theta_{k}$ with
\begin{subequations}
\begin{alignat}{2}
\sin\left(2\theta_{k}\right)&=\frac{J_{0}\sin k}{\sqrt{J_{0}^{2}+B_{x}^{2}-2B_{x}J_{0}\cos k}} \label{eq:theta1}\\
\cos\left(2\theta_{k}\right)&=\frac{B_{x}-J_{0}\cos k}{\sqrt{J_{0}^{2}+B_{x}^{2}-2B_{x}J_{0}\cos k}}. \label{eq:theta2}
\end{alignat}
\end{subequations}
In terms of the quasiparticle modes, the Hamiltonian (\ref{eq:cfermion-k}) becomes
\begin{equation}
H_{s}=\sum_{k}\varepsilon_{k}\hat{\gamma}_{k}^{\dag}\hat{\gamma}_{k}+E_{g},\label{eq:H0k}
\end{equation}
where the summation is taken over all wave vectors (both $k\ge0$ and $k<0$). The eigenenergy of quasiparticle $\hat{\gamma}_{k}$ is 
\begin{equation}
\varepsilon_{k}=2\sqrt{J_{0}^{2}+B_{x}^{2}-2B_{x}J_{0}\cos k}\label{eq:omK}
\end{equation}
with $\varepsilon_{k}=\varepsilon_{-k}$. At the quantum critical point $B_{x}=J_{0}$, $\varepsilon_{k}=4J_{0}|\sin(k/2)|$, with $\varepsilon_{k}\rightarrow 0$ as $k\rightarrow 0$. The ground-state energy of the TFIM is $E_{g}=\sum_{k>0}\varepsilon_{k}[\cos(2\theta_{k})-1]-NB_{x}$. Equation (\ref{eq:H0k}) shows that the TFIM is an integrable model, where quasiparticles in different $(k,-k)$ subspace are not coupled.

\subsection{Ground state properties\label{ssec:ground}}
The ground state of the TFIM satisfies $H_{s}\vert G\rangle=E_{g}\vert G\rangle$. For an arbitrary quasimomentum $k$, we have $\hat{\gamma}_{k}\vert G\rangle=0$. In its most general form, the ground state can be written as
\begin{equation}
\left|G\right\rangle =\prod_{k>0}(a_{k}+b_{k}\hat{c}_{k}^{\dag}+b_{-k}\hat{c}_{-k}^{\dag}+d_{k}\hat{c}_{k}^{\dag}\hat{c}_{-k}^{\dag})\vert0_{k,-k}\rangle,
\end{equation}
which contains superposition of all possible states in each $(k,-k)$ subspace. Here $\vert 0_{k,-k}\rangle$ is the vacuum state for the $(k,-k)$ modes. Using the conditions $\hat{\gamma}_{k}\vert G\rangle=0$ and $\hat{\gamma}_{-k}\vert G\rangle=0$, we find 
\begin{equation}
b_{k}=0,\,\, b_{-k}=0,\,\,d_{k}u_{k}=\textrm{i}v_{k}a_{k}. \label{eq:bkdk}
\end{equation}
Combining these results with the normalization condition, the ground-state wave function can be written as $\vert G\rangle=\prod_{k>0}\vert G_{k}\rangle$, with
\begin{equation}
\vert G_{k}\rangle=\left(u_{k}+\textrm{i}v_{k}\hat{c}_{k}^{\dag}\hat{c}_{-k}^{\dag}\right)\left\vert0_{k,-k}\right\rangle\label{eq:gnd}
\end{equation}
being the ground state in the $(k,-k)$ subspace. It can also be shown that 
\begin{subequations}
\begin{alignat}{2}
\hat{\gamma}_{k}^{\dag}\vert G_{k}\rangle&=\hat{c}_{k}^{\dag}\vert0_{k,-k}\rangle,\label{eq:gammakG1}\\
\hat{\gamma}_{-k}^{\dag}\vert G_{k}\rangle&=\hat{c}_{-k}^{\dag}\vert0_{k,-k}\rangle,\label{eq:gammakG2}
\end{alignat}
\end{subequations}
which tell us that the creation of one quasiparticle $\hat{\gamma}_{\pm k}$ from the ground state $\vert G_{k}\rangle$ corresponds to the generation of one physical particle $\hat{c}_{\pm k}$ from the vacuum state in the physical basis. Furthermore,
\begin{equation}
\textrm{i}\hat{\gamma}_{k}^{\dag}\hat{\gamma}_{-k}^{\dag}\vert G_{k}\rangle=\left(-v_{k}+\textrm{i}u_{k}\hat{c}_{k}^{\dag}\hat{c}_{-k}^{\dag}\right)\vert0_{k,-k}\rangle,\label{eq:ggkgnd}
\end{equation}
which contains two quasiparticles, one in the $\hat{\gamma}_{k}$ mode and the other in the $\hat{\gamma}_{-k}$ mode. This state is orthogonal to the ground state $\vert G_{k}\rangle$ in the $(k,-k)$ subspace. 

\subsection{Time-dependent wave function\label{ssec:timedependent}}
The Hamiltonian (\ref{eq:cfermion-k}) (and similarly, $\widetilde{H}_{s}$ under effective transverse field $\widetilde{B}_{x}$ in the cavity-coupled model) only contains the following terms: $\hat{c}_{k}^{\dagger}\hat{c}_{k}$, $\hat{c}_{k}^{\dagger}\hat{c}_{-k}^{\dag}$ and $\hat{c}_{-k}\hat{c}_{k}$. These terms conserve the parity of the states of the TFIM during arbitrary time evolution, i.e., the number of fermionic particles will remain in the even (or odd) subspace during the entire evolution. Hence starting from an initial state with even (odd) number of fermionic particles, the TFIM governed by a time-dependent Hamiltonian will always have even (odd) number of particles during the evolution. In particular, starting from the ground state of an initial (effective) transverse field, the number of fermionic particles or quasiparticles will always be an even number as the (effective) transverse field varies with time. The wave function at arbitrary time $t$ can hence always be written as $|\psi(t)\rangle =\prod_{k>0}(U_{k}(t)+\textrm{i}V_{k}(t)\hat{c}_{k}^{\dagger}\hat{c}_{-k}^{\dag})|0\rangle$ with time-dependent coefficients $U_{k}(t)$ and $V_{k}(t)$. 

Assume that the instantaneous Hamiltonian at time $t$ is $H_{s}(t)$ with ground state $\vert G(t)\rangle=\prod_{k>0}\vert G_{k}(t)\rangle$. The above wave function can also be decomposed as
\begin{equation}
|\psi(t)\rangle=\prod_{k>0}[\alpha_{k}(t)+\textrm{i}\beta_{k}(t)\hat{\gamma}_{k}^{\dag}\hat{\gamma}_{-k}^{\dag}]|G_{k}(t)\rangle,\label{eq:psiab}
\end{equation}
where $\hat{\gamma}_{k}^{\dag}$ is the creation operator for the instantaneous quasiparticle mode with quasimomentum $k$. The probability amplitudes $\alpha_{k}$ and $\beta_{k}$ can be derived as 
\begin{subequations}
\begin{alignat}{2}
\alpha_{k}(t)&=[u_{k},v_{k}]\cdot [U_{k}(t), V_{k}(t)]^{\textrm{T}},\label{eq:alpha} \\
\beta_{k}(t)&=[-v_{k},u_{k}]\cdot [U_{k}(t), V_{k}(t)]^{\textrm{T}},\label{eq:beta}
\end{alignat}
\end{subequations}
with ``T'' referring to the transpose operation. The wave function in the $(k,-k)$ subspace is then a superposition of the ground state $|G_{k}(t)\rangle$ with probability amplitude $\alpha_{k}(t)$ and the excited state $\textrm{i}\hat{\gamma}_{k}^{\dag}\hat{\gamma}_{-k}^{\dag}|G_{k}(t)\rangle$ with probability amplitude $\beta_{k}(t)$. 

\subsection{Average of the operator $\sum_{i}\hat{\sigma}_{xi}$\label{ssec:Xave}}
When the wave function of the TFIM is known, the averages of various quantum operators can be calculated directly from the wave function. From these averages, properties of the TFIM can be inferred. For example, the equal-time correlation function $C_{ij}=\langle G|\hat{\sigma}_{zi}\hat{\sigma}_{zj}|G\rangle$ between the operators $\hat{\sigma}_{zi}$ and $\hat{\sigma}_{zj}$ can be used to characterize quantum phase transition in the TFIM. In our system, the cavity mode is coupled to the TFIM via the operator $\sum_{i}\hat{\sigma}_{xi}$. The average of this operator plays an important role in the dynamics of this system. 

In the representation of the fermionic modes, we can write
\begin{equation}
\sum_{i}\hat{\sigma}_{xi} = N- 2\sum_{k}\hat{c}_{k}^{\dagger}\hat{c}_{k}.\label{eq:sx_ck}
\end{equation}
For an arbitrary state $\vert \psi(t)\rangle=\prod_{k>0}(U_{k}+\textrm{i}V_{k}\hat{c}_{k}^{\dag}\hat{c}_{-k}^{\dag})\vert0\rangle$, we derive the average of the $\sum_{i}\hat{\sigma}_{xi}$ operator as
\begin{equation}
X=\langle \psi(t)| \sum_{i}\hat{\sigma}_{xi} |\psi(t)\rangle = N-4 \sum_{k>0}|V_{k}|^{2},\label{eq:Xavepsi}
\end{equation}
which is determined by the coefficients $\{ V_{k}\}$. For the ground state $|G\rangle$ of a large lattice ($N\rightarrow\infty$), we can convert the summation into an integration, with
\begin{eqnarray}
X&=&\langle G | \sum_{i}\hat{\sigma}_{xi} |G\rangle \nonumber \\ 
&=& \frac{N}{2\pi}\int_{-\pi}^{\pi}dk\frac{B_{x}-J_{0}\cos k}{\sqrt{J_{0}^{2}+B_{x}^{2}-2B_{x}J_{0}\cos k}}.\label{eq:Xave}
\end{eqnarray}
In the limit of $B_{x}\gg J_{0}$, $\langle G|\sum_{i}\hat{\sigma}_{xi}|G\rangle\rightarrow N$; and for $B_{x}\ll J_{0}$, $\langle G|\sum_{i}\hat{\sigma}_{xi}|G\rangle\rightarrow0$.

For the cavity-coupled TFIM studied in this work, the stationary state of the TFIM is the ground state under the effective transverse field $\widetilde{B}_{x}=B_{x}-gx_{a}^{(s)}$, which contains a shift induced by the stationary cavity displacement $x_{a}^{(s)}$. The stationary-state average $X^{(s)}$ of the $\sum_{i}\hat{\sigma}_{xi}$ operator is hence the ground-state average of this operator under the effective transverse field $\widetilde{B}_{x}$, which depends on the stationary cavity displacement $x_{a}^{(s)}$. The stationary values of $X^{(s)}$ and $x_{a}^{(s)}$ can be solved self-consistently.

\section{Bifurcation, stability condition, and phase switchings\label{sec:stability}}
In this section, we present derivations of the stationary solutions, stability condition, bifurcation points and phase switchings at these points.

\subsection{Stationary solutions\label{ssec:ss}}
Our system contains a TFIM coupled to a cavity mode. For the cavity, the Langevin equation for the average of the annihilation operator is given by Eq.~(\ref{eq:dat}) in Sec.~\ref{sec:bifurcation}. The conjugate equation for Eq.~(\ref{eq:dat}) is 
\begin{equation}
\frac{d}{dt}\left\langle \hat{a}^{\dag}\right\rangle  =- \textrm{i}\Delta_{c}\left\langle \hat{a}^{\dag}\right\rangle -\frac{\kappa}{2}\left\langle \hat{a}^{\dag}\right\rangle -\textrm{i}\left(\epsilon-gX\right),\label{eq:dadagt}
\end{equation}
where we assume that the driving amplitude is a real number for simplicity of discussion. These equations depend on the average $X$ of the $\sum_{i}\hat{\sigma}_{xi}$ operator, and hence depend on the state of the TFIM.

\begin{figure}
\includegraphics[clip,width=7cm]{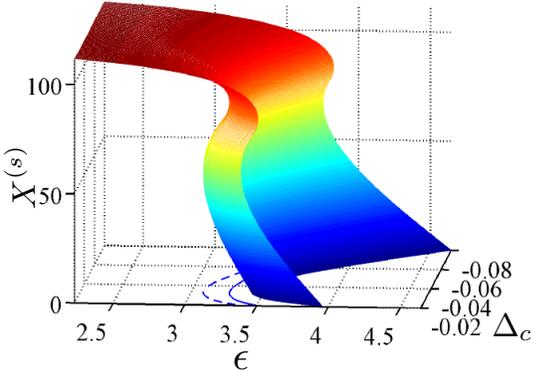}
\caption{(Color online) The stationary-state average $X^{(s)}$ for the operator $\sum_{i}\hat{\sigma}_{xi}$ vs. driving amplitude $\epsilon$ and cavity detuning $\Delta_{c}$. Solid (dashed) line: the bifurcation point $\epsilon_{2}$ ($\epsilon_{1}$) vs. $\Delta_{c}$. The parameters are the same as these in Fig.~\ref{fig1}(b).}
\label{figS1} 
\end{figure}

Let $x_{a}=\langle (\hat{a}+\hat{a}^{\dag})\rangle$ and $p_{a}=-\textrm{i}\langle (\hat{a}-\hat{a}^{\dag})\rangle $ be the averages of the coordinate and momentum quadratures of the cavity mode, respectively. From Eqns.~(\ref{eq:dat}) and (\ref{eq:dadagt}), we have
\begin{subequations}
\begin{alignat}{2}
\frac{d}{dt}x_{a} & =  -\Delta_{c} p_{a}-\frac{\kappa}{2}x_{a},\label{eq:dxa}\\
\frac{d}{dt}p_{a} & =  \Delta_{c} x_{a}-\frac{\kappa}{2}p_{a}+2\left(\epsilon-gX\right).\label{eq:dpa}
\end{alignat}
\end{subequations}
Stationary solutions of these equations satisfy the relations $dx_{a}/dt=0$ and $dp_{a}/dt=0$. The stationary state of the TFIM is the ground state under the effective transverse field $\widetilde{B}_{x} = B_{x}-gx_{a}^{(s)}$, which is shifted by the nonzero cavity displacement $x_{a}^{(s)}$. The average of the $\sum_{i}\hat{\sigma}_{xi}$ operator in the stationary state is then $X^{(s)}=\langle G|\sum_{i}\hat{\sigma}_{xi}|G\rangle$ with $|G\rangle$ being the ground state of the TFIM under $\widetilde{B}_{x}$. As the effective transverse field $\widetilde{B}_{x}$ is directly related to $x_{a}^{(s)}$, the operator average $X^{(s)}$ depends on $x_{a}^{(s)}$ as well. We obtain
\begin{subequations}
\begin{alignat}{2}
x_{a}^{(s)} &= \frac{2\Delta_{c}\left[-\epsilon+gX^{(s)}\right]}{\Delta_{c}^{2}+\kappa^{2}/4},\label{eq:avexa} \\
p_{a}^{(s)} &= -\left(\kappa/2\Delta_{c}\right)x_{a}^{(s)}.\label{eq:avepa}
\end{alignat}
\end{subequations}
The operator average $X^{(s)}$ has a nonlinear dependence on the stationary cavity displacement $x_{a}^{(s)}$. This nonlinear dependence results in a bistable regime in this coupled system. In Fig.~\ref{figS1}, we plot $X^{(s)}$ as a function of $\epsilon$ and $\Delta_{c}$, which clearly demonstrates the bistability of the cavity-coupled TFIM. Similar results are also given in Fig.~\ref{fig1}, where the stationary cavity displacement $x_{a}^{(s)}$ is plotted vs. driving amplitude $\epsilon$ and detuning $\Delta_{c}$. 

In the bistable regime, there exist three stationary solutions at a given driving amplitude $\epsilon$. The solution with the smallest $x_{a}^{(s)}$ corresponds to $\widetilde{B}_{x}>1$, with the TFIM deep within the paramagnetic phase. The solution with the largest $x_{a}^{(s)}$ has $\widetilde{B}_{x}<1$ with the TFIM in the ferromagnetic phase. The third solution has an intermediate value of $x_{a}^{(s)}$ in-between the first two solutions.  

\subsection{Stability condition\label{ssec:stability}}
Here we study the stability of the stationary solutions. Assume that the cavity has a small offset $\widetilde{x}_{a}$ ($\widetilde{p}_{a}$) from its stationary value of the coordinate (momentum) quadrature, i.e., $x_{a}= x_{a}^{(s)}+\widetilde{x}_{a}$ ($p_{a}=p_{a}^{(s)}+\widetilde{p}_{a}$). We linearize the operator average $X$ in terms of these offsets as
\begin{equation}
X=  X^{(s)}(x_{a}^{(s)})+X^{\prime}\cdot\widetilde{x}_{a}\label{eq:X'}
\end{equation}
with the derivative $X^{\prime}=dX^{(s)}/dx_{a}^{(s)}$. Using Eqns.~(\ref{eq:dxa}, \ref{eq:dpa}, \ref{eq:X'}), we obtain
\begin{equation}
\frac{d}{dt}\left[\begin{array}{c}
\widetilde{x}_{a}\\
\widetilde{p}_{a}
\end{array}\right]=\left[\begin{array}{cc}
-\kappa/2 & -\Delta_{c}\\
\Delta_{c}-2gX^{\prime} & -\kappa/2
\end{array}\right]\left[\begin{array}{c}
\widetilde{x}_{a}\\
\widetilde{p}_{a}
\end{array}\right],\label{eq:ddx_ddp}
\end{equation}
which are dynamic equations for these small offsets. Let the eigensolutions of Eq.~(\ref{eq:ddx_ddp}) have the time dependence $e^{\lambda t}$. The secular frequencies can be derived as
\begin{equation}
\lambda_{\pm}=-\frac{\kappa}{2}\pm\sqrt{\left(2\Delta_{c} gX^{\prime}-\Delta_{c}^{2}\right)}.\label{eq:om12}
\end{equation}
When the real part of one of the secular frequencies is positive, the stationary solution becomes unstable~\cite{DeJesusPRA1987}. The stability condition of this system is then
\begin{equation}
\Delta_{c}<0 \quad \textrm{and}\quad \frac{\Delta_{c}^{2}+\kappa^{2}/4}{2\Delta_{c}}<gX^{\prime},\label{eq:stability}
\end{equation}
which requires a negative cavity detuning. 

An interesting property of the stability condition is its connection to the derivative of the driving amplitude $\epsilon$ with respect to the stationary cavity displacement $x_{a}^{(s)}$. By conducting a differentiation with respect to $x_{a}^{(s)}$ on both sides of Eq.~(\ref{eq:avexa}), we have
\begin{equation}
\frac{d\epsilon}{dx_{a}^{(s)}}=gX^{\prime}-\frac{\Delta_{c}^{2}+\kappa^{2}/4}{2\Delta_{c}}.\label{eq:depdxa}
\end{equation}
Hence, the stability of a stationary solution (\ref{eq:stability}) is equivalent to the positivity of its slope, i.e., $d\epsilon/dx_{a}^{(s)}>0$. Among the three solutions at a given driving amplitude, the solution with the smallest and the largest $x_{a}^{(s)}$ are both stable with positive slopes. The solution with the smallest (largest) $x_{a}^{(s)}$ corresponds to the TFIM deep within the paramagnetic (ferromagnetic) phase. The solution with the intermediate $x_{a}^{(s)}$ is unstable with negative slope $d\epsilon/dx_{a}^{(s)}<0$. Outside the bistable regime, the system has only one stable solution at a given $\epsilon$, as shown in Fig.~\ref{fig1}.

\subsection{Bifurcation points and sudden phase switchings\label{ssec:bifurcation}}
In our system, the bifurcation points $\epsilon_{1,2}$ separate the bistable regime from regimes with only one stationary solution, and satisfy the condition 
\begin{equation}
d\epsilon/dx_{a}^{(s)}=0.\label{eq:bifurpoint}
\end{equation}
With Eq.~(\ref{eq:depdxa}), this condition can be expressed as
\begin{equation}
gX^{\prime}=\frac{\Delta_{c}^{2}+\kappa^{2}/4}{2\Delta_{c}}.\label{eq:bpoint}
\end{equation}
At the bifurcation points, the stationary phase of the TFIM changes abruptly from paramagnetic to ferromagnetic, or vice versa. Such switchings involve finite changes in system energy, effective transverse field $\widetilde{B}_{x}$, and stationary cavity displacement $x_{a}^{(s)}$, and resemble a first-order phase transition, as studied in detail in \cite{TianPRL2010}. The dynamical quantum phase transition and how the switchings occur across the bifurcation points are studied in this work. Below we discuss several interesting properties of the bifurcation points in the cavity-coupled TFIM under the coupling term $g(\hat{a}+\hat{a}^{\dag})\sum_{i}\hat{\sigma}_{xi}$.

\begin{figure}
\includegraphics[clip,width=\columnwidth]{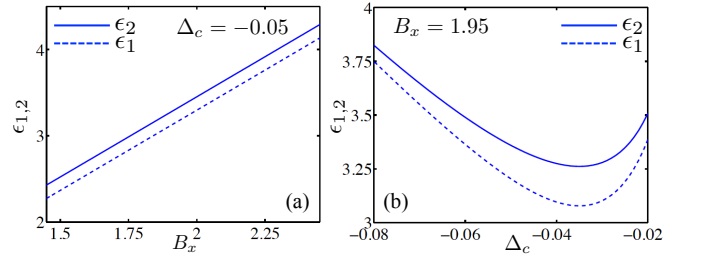}
\caption{(Color online) (a) The bifurcation points $\epsilon_{1,2}$ vs. transverse field $B_{x}$ at a given detuning. (b) The bifurcation points $\epsilon_{1,2}$ vs. detuning $\Delta_{c}$ at a given transverse field. Solid lines: $\epsilon_{2}$; dashed lines: $\epsilon_{1}$. The parameters are the same as these in Fig.~\ref{fig1}(b).}
\label{figS2} 
\end{figure}
\begin{figure}
\includegraphics[clip,width=\columnwidth]{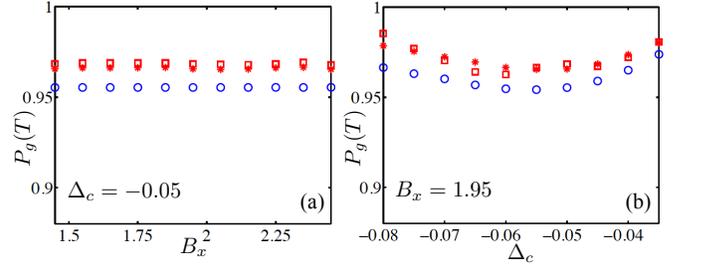}
\caption{(Color online) (a) and (b) The probability $P_{g}(T)$ vs. $B_{x}$ and vs. $\Delta_{c}$ at the final time $T=400$, respectively. Blue circles: numerical results for the cavity-coupled TFIM; red squares: numerical results for the linearly-ramped TFIM; red stars: results from the Landau-Zener formula for the linearly-ramped TFIM. The parameters are the same as these in Fig.~\ref{fig1}(b).}
\label{figS3} 
\end{figure}
From Eq.~(\ref{eq:bpoint}), we see that the derivative $X^{\prime}$ at the bifurcation points does not depend on the original transverse field $B_{x}$ in the TFIM. This observation indicates that the effective transverse field $\widetilde{B}_{x}$ at the bifurcation points, as well as the operator average $X^{(s)}$, does not depend on $B_{x}$ either. Because $\widetilde{B}_{x}=B_{x}-gx_{a}^{(s)}$, the stationary cavity displacement $x_{a}^{(s)}$ at the bifurcation points thus has a linear dependence on the original transverse field $B_{x}$. Combining these results with Eq.~(\ref{eq:avexa}), we find that the driving amplitudes ($\epsilon_{1,2}$) at the bifurcation points also have a linear dependence on $B_{x}$ at given values of $\Delta_{c}$, $\kappa$, and $g$. This is clearly illustrated in Fig.~\ref{figS2}(a), where $\epsilon_{1,2}$ are plotted as functions of $B_{x}$ with all other parameters fixed. Meanwhile, as $\widetilde{B}_{x}$ is independent of $B_{x}$, the states of the TFIM at the bifurcation points and the dynamics of the TFIM after a sudden quench at these points do not depend on $B_{x}$ either. This argument is confirmed by our numeral simulation of the time evolutions of the cavity-coupled TFIM at different values of $B_{x}$. As shown in Fig.~\ref{figS3}(a), the probability $P_{g}(t=T)$ of the ground state at the final time $T=400$ is almost independent of $B_{x}$, with only small variations resulted from numerical inaccuracies. 

In contrast, the derivative $X^{\prime}$ at the bifurcation points $\epsilon_{1,2}$ depends on the detuning $\Delta_{c}$. As a result, the effective transverse field $\widetilde{B}_{x}$ at $\epsilon_{1,2}$ varies with the detuning, which affects the dynamics and the quasiparticle excitations during the switching processes. The bifurcation points $\epsilon_{1,2}$ thus have more complicated dependence on the detuning $\Delta_{c}$ than on $B_{x}$. This is shown in Fig.~\ref{figS2}(b), where $\epsilon_{1,2}$ are plotted as functions of the detuning $\Delta_{c}$. To illustrate this dependence, the probability $P_{g}(t=T)$ of the ground state at the end of a dynamical evolution at time $T=400$ is plotted versus the detuning in Fig.~\ref{figS3}(b).

\end{document}